# New class of *T'*-structure cuprate superconductors

A. Tsukada[1], Y. Krockenberger[1,2], M. Noda[1,3], H. Yamamoto[1], D. Manske[2], L. Alff[4], and M. Naito[1,5]

[1]*NTT Basic Research Laboratories, 3-1 Morinosato, Atsugi, Kanagawa 243, Japan*

[2]*Max-Planck-Institute for Solid State Research, Heisenbergstr. 1, 70569 Stuttgart, Germany*

[3]*Tokyo University of Science, 2641 Yamazaki, Noda, Chiba 278-8510, Japan*

[4]*Vienna University of Technology, ISAS, Applied Electronic Materials, Gusshausstr. 27-29/366, A-1040 Wien, Austria*

[5]*Department of Applied Physics, Tokyo University of Agriculture and Technology (TUAT), 2-24-16 Naka-cho, Koganei, Tokyo 184-8588, Japan*

**High-temperature superconductivity has been discovered in $La_{2-x}Ba_xCuO_4$ [1], a compound that derives from the undoped $La_2CuO_4$ crystallizing in the perovskite *T*-structure. In this structure oxygen octahedra surround the copper ions. It is common knowledge that charge carriers induced by doping in such an undoped antiferromagnetic Mott-insulator lead to high-temperature superconductivity [2-4]. The undoped material $La_2CuO_4$ is also the basis of the electron-doped cuprate superconductors [5] of the form $La_{2-x}Ce_xCuO_{4+y}$ [6,7] which however crystallize in the so called *T'*-structure, i.e. without apical oxygen above or below the copper ions of the $CuO_2$-plane. It is well known that for $La_{2-x}Ce_xCuO_{4+y}$ the undoped *T'*-structure parent compound *cannot* be prepared due to the structural phase transition back into the *T*-structure occuring around $x \sim 0.05$. Here, we report that if La is substituted by *RE* = Y, Lu, Sm, Eu, Gd, or Tb, which have smaller ionic** **radii but have the same valence as La, *nominally undoped* $La_{2-x}RE_xCuO_4$ can be synthesized by molecular beam epitaxy in the *T'*-structure. The second important result is that all these new *T'*-compounds are *superconductors* with fairly high critical temperatures up to 21 K. For this new class of cuprates $La_{2-x}RE_xCuO_4$, which forms the T'-parent compounds of the La-based electron doped cuprates, we have not been able to obtain the Mott-insulating ground state for small *x* before the structural phase transition into the *T*-structure takes place.**

For the hole-doped high-$T_C$ superconductors $La_{2-x}RE_xCuO_4$ (with $RE$ = Ca, Sr, Ba) the highest $T_C$ occurs at a doping level of about 0.15. This is also the case for the electron-doped high-$T_C$ superconductors $LN_{2-x}Ce_xCuO_4$ (with $LN$ = Pr or Nd) where electrons are induced by doping of Ce atoms which have valency 4+ in the $Nd_2CuO_4$-structure. In contrast to the hole-doped compounds however, the phase diagram of these electron doped high-$T_C$ superconductors displays a much larger antiferromagnetic region and a much smaller superconducting region [8,9]. In a simple physical picture this is due to a spin dilution effect created by electron doping eliminating copper spins without creation of spin frustration as in the hole doped systems [10]. However, we find that the phase diagram of the electron-doped high-$T_C$ superconductors strongly depends on the *size* of the lanthanide (*LN*) ion. To be more precise, the superconducting range increases with increasing ionic radius of the *LN*-ion, whereas the corresponding antiferromagnetic region decreases (see Figure 1). For the largest lanthanide which is La itself, the maximal $T_C$ of about 30 K occurs at a doping level around 0.1 [6,7]. A second interesting feature in the phase diagram of the electron-doped high-$T_C$ superconductors is the *sudden* disappearance [8,11] of superconductivity towards lower doping (see Figure 1). By improving the oxygen removal process in the compound $Pr_{2-x}Ce_xCuO_{4+y}$, the superconducting region can be extended even with slightly increasing critical temperature down to a doping level of around $x$ = 0.05 [12]. In the case of $La_{2-x}Ce_xCuO_{4+y}$ it is the structural phase transition around a doping level of 0.05 into the *T*- structure that leads to strongly insulating behavior for $x < 0.05$ [6]. The absence of the well-known 'dome'-shaped superconducting region of the hole-doped high-$T_C$ cuprates in their electron-doped counterparts, and the abrupt disappearance of superconductivity at low doping raises the question concerning the nature of the undoped *T'*-structure $La_2CuO_4$.

Using molecular beam epitaxy at low temperatures it is possible to stabilize the undoped *T'*-structure of $La_2CuO_4$ as thin film on a $SrTiO_3$ substrate [13], however with lower crystallinity as compared to *T*-structure $La_2CuO_4$. The interesting finding is, that the different crystal structure of the $La_2CuO_{4\pm y}$ thin films with $y \sim 0$ results in a difference of resistivity of four to five orders of magnitude: While $La_2CuO_4$ in the *T*-structure is strongly insulating, the same compound in the *T'*-structure is almost metallic (with $d\rho/dT \geq 0$) down to 150 K with increasing resistivity (with $d\rho/dT < 0$) at low temperatures (see Figure 2). The metallicity of *T'*-structure $La_2CuO_4$ is in agreement with the resistivity evolution of the series $LN_2CuO_4$ with *increasing* ionic radius of the *LN*-ion (*LN* = Tb, Gd, Eu, Sm, Nd, Pr, La): The increased ionic radius results evidently in a more and more metallic behavior (see Figure 3).

A simple trick allows the fabrication of a *new class* of undoped *T'*-structure compounds with good crystallinity by thin film synthesis using molecular beam epitaxy. Most films were grown at relatively low substrate temperatures around 650°C on $SrTiO_3$ substrates. The replacement of La by smaller but *iso*valent ions such as Y or Tb stabilizes the *T'*-structure, a fact which confirms that the phase transition is governed by the ionic size of the lanthanide ion. Upon substituting $La^{3+}$ by $Tb^{3+}$ (and similarly other lanthanides with smaller radius than $La^{3+}$) the phase transition into the *T'*-structure takes place at a Tb concentration of about 5%. The thin films of the composition $La_{2-x}Tb_xCuO_{4+y}$ are stable and single phase within a wide range of $x$ (see Figure 4). This allows for the first time the study of the La-based *T'*-structure $La_{2-x}RE_xCuO_{4+y}$ *without* doping-induced charge carriers, i.e. the material from which the family of the electron doped high-temperature superconductors is derived. The 3+ state of Y and Tb has been clearly established by x-ray photoelectron spectroscopy (see Figure 5). Identical behavior is found for $Y^{3+}$, $Lu^{3+}$, $Sm^{3+}$, $Eu^{3+}$, $Gd^{3+}$, and $Tb^{3+}$.

The main result of our study is that this new class of cuprate compounds with *T'*-structure *without charge carriers induced by non-isovalent dopants* is metallic ($d\rho/dT > 0$ for $T > 75$ K), and show superconductivity at critical temperatures up to 21 K. The superconducting state is clearly evidenced by resistivity measurements, and by flux expulsion due to the Meißner-Ochsenfeld-effect as observed by magnetometry (see Figure 6). For thin films it is not possible to estimate the Meissner fraction, however, the observation of fairly high critical currents (above $10^5$ A/cm²) clearly confirms a high Meissner fraction in the thin films and excludes filamentary superconductivity.

How to establish that the investigated *T'*-structure materials indeed have no doping-induced charge carriers? It is safe to exclude higher *valencies* of the *LN*$^{3+}$ ions as this is clearly established by x-ray photoelectron spectroscopy (see Fig. 5). Also one would not expect that for example $Y^{3+}$ has a deviating valency. Furthermore, *strain effects* [14] can be excluded, since the thin films in our studies are up to 200 nm thick. What remains is the oxygen content of the samples, where *y* represents the deviation from the desired $O_4$ status. Unfortunately, with present technologies it is impossible to determine directly the oxygen content of a thin film (on an oxide substrate). While *excess oxygen* does lead to metallicity (or superconductivity) in the hole doped cuprates, this is not the case for electron-doped cuprates. Since the thin films are clearly in the *T'*-structure, a small amount of residual hole-doped copper-oxygen planes with local *T*-structure cannot account for the superconductivity with high $T_C$ and high critical currents as observed here. In order to obtain superconductivity in electron-doped high-$T_C$ superconductors, the samples have to be *oxygen reduced*. From single crystal thermogravimetric analysis in $Nd_{2-x}Ce_xCuO_{4+y}$ it is known that the amount of oxygen removed by the reduction process is of the order $\Delta y \sim 0.02$ [15]. Due to the enhanced surface to volume ratio, it is very likely that the removal process is more complete for thin films. Neutron diffraction on $Nd_{2-x}Ce_xCuO_{4+y}$ single crystals has revealed that during the removal process mostly interstitial (apical) oxygen (O(3)-position) is eliminated, while the oxygen ions in the copper oxide plane (O(1)-position) and the out-of-plane oxygen ions (O(2)-position) are not affected within the resolution of the experiment [16,17]. This demonstrates that the oxygen binding energy at the O(3) site is clearly lower than those at the O(1) and O(2) sites. For the undoped *T*-structure $La_2CuO_{4+y}$, oxygen or ozone annealing can induce superconductivity by excess oxygen ($\Delta y \sim 0.1$ [18]). By this excess oxygen, holes are induced in the same way as by the substitution of $Sr^{2+}$ for $La^{3+}$ [19]. Our oxygen reduction studies for *T'*-structure $La_{1.85}Y_{0.15}CuO_{4+y}$, however, show that it is not possible to reach an insulating state close to $y \sim 0$ (see Fig. 7). Samples cooled in the highest oxygen pressure are insulating. Since similar oxygen pressures in the case of the *T*-structure lead to a large amount of excess oxygen [19], we assume high oxygen pressures correspond to $y > 0$. If the samples are exposed to atmospheres with decreasing oxygen partial pressure atmospheres, therefore, $y$ is approaching zero from above. At the same time, the resistivity drops *continuously* and eventually superconductivity occurs. There is no trace of an insulating state close to $y \sim 0$. Further reduction leads to defects, since removal of O(1) or O(2) oxygen destabilizes the crystal structure (secondary phases like $(La,Nd,Ce)_2O_3$ may form [20,21]). These defects cannot be reversed by oxygen annealing. The appearance of the defects can immediately be seen in-situ by reflection high-energy electron diffraction. In short, while we cannot rule out a small *oxygen deficiency* in the sample, the remarkable point is, that there is no sign for the (expected) insulating phase around $y \sim 0$. We thus can conclude that we have investigated $La_{2-x}RE_xCuO_{4+y}$ samples with $y$ close to zero. The important result is that for these compounds, instead of an insulating behavior, metallicity and even high-$T_C$ superconductivity is found.

It is commonly believed that high-$T_C$ superconductivity is closely related to strong electron correlations leading to the formation of an upper and (effective) lower Hubbard band with the Fermi energy lying in the undoped case in the gap between the two bands [2-4,22,23]. However, the fundamental problem of the transition from the Mott-insulating state into the metallic/superconducting state is still not well understood [4,24-28]. Our systematic experiments suggest that this transition from an antiferromagnetic Mott-insulating state into the metallic regime is affected by the crystal structure i.e. by the size of the ionic radii of the involved rare earth ions. For the newly synthesized La-based cuprate compounds of the form $La_{2-x}RE_xCuO_{4+y}$ with stabilized *T'*-structure reported here, the Mott-insulating ground state is not obtained at all.

Figure 1: Phase diagram of electron-doped high $T_C$–superconductors $LN_{2-x}Ce_xCuO_4$ with $LN$ = Eu, Sm, Nd, Pr, and La. The width of the superconducting region increases, and the maximum $T_{C,0}$ (where the resistance vanishes) shifts to lower doping values with increasing rare earth ion radius (for eigth-fold coordination: $Eu^{3+}$: 1.066 Å, $Sm^{3+}$: 1.079 Å, $Nd^{3+}$: 1.109 Å, $Pr^{3+}$: 1.126 Å, $La^{3+}$: 1.160 Å). Note that the superconducting region of $La_{2-x}Ce_xCuO_4$ spans the whole superconducting region of all other compounds. At the low doping side, superconductivity disappears *abruptly* (in contrast to the phase diagram of the hole-doped cuprates with its famous 'dome'-shaped superconducting region). The antiferromagnetic region at low doping is not shown in the diagram, since the Néel-temperatures for the thin films have not been determined.

Figure 2: Comparison of resistivity of $La_2CuO_{4\pm y}$ thin films with the *T*- and *T'*– structure for $y \sim 0$ [14]. The different crystal structures are sketched. The structural phase transition is coupled to a change in resistivity of several orders of magnitude. *T'* - structure $La_2CuO_{4\pm y}$ thin films are metallic down to about 180 K ($d\rho/dT > 0$) while *T* - structure $La_2CuO_{4\pm y}$ thin films are insulating ($d\rho/dT < 0$). This sample could not be measured in our experimental set-up below 150 K due to the high resistivity.

Figure 3: Resistivity of the series $LN_2CuO_{4+y}$ with $LN$ = Tb, Gd, Eu, Sm, Nd, Pr, and La (from above to below). All materials have *T'* – structure. The resistivity decreases with increasing ionic radius of the lanthanide ions within the series (cf. Figure 1, $Tb^{3+}$: 1.040 Å, $Gd^{3+}$: 1.053 Å).

Figure 4: Out-of-plane lattice parameter $c$ vs. Tb doping $x$ in $La_{2-x}Tb_xCuO_{4+y}$. Due to the smaller size of the $Tb^{3+}$ ion compared to $La^{3+}$ the $T'$ - structure is stabilized within a wide range of doping $x$. The typical FWHM of the x-ray diffraction rocking curve is about 0.1°. The crystal structure is sketched: In the $T$ – structure the copper ion is six-fold and the La ion nine-fold coordinated; in the $T'$ – structure the copper ion is four-fold and the La ion eight-fold coordinated

Figure 5: In-situ Tb-4$d$ x-ray photo emission (XPS) spectrum of $La_{1.7}Tb_{0.3}CuO_4$ thin film. For comparison the corresponding spectra of $TbO_2$ ($Tb^{4+}$), $Tb_4O_7$ (mixed valence $Tb^{3+}/Tb^{4+}$), and $Tb_2O_3$ ($Tb^{3+}$) are shown [29].

Figure 6: Superconductivity in $La_{2-x}Y_xCuO_{4+y}$ ($T_{C,0}$ ~ 21 K) and $La_{2-x}Tb_xCuO_{4+y}$ ($T_{C,0}$ ~ 12 K) from resistivity and magnetometry measurements respectively. The Meißner-Ochsenfeld effect was measured using a superconducting quantum interference device.

Figure 7: Oxygen reduction study for $La_{1.85}Y_{0.15}CuO_{4+y}$. For each measured sample, the oxygen/ozone pressure atmosphere is indicated. For the highest oxygen pressure we estimate $y$ ~ 0.1. With increasingly reducing atmosphere the resistivity drops continuously (even after superconductivity is obtained). Further reduction results in sample decomposition.

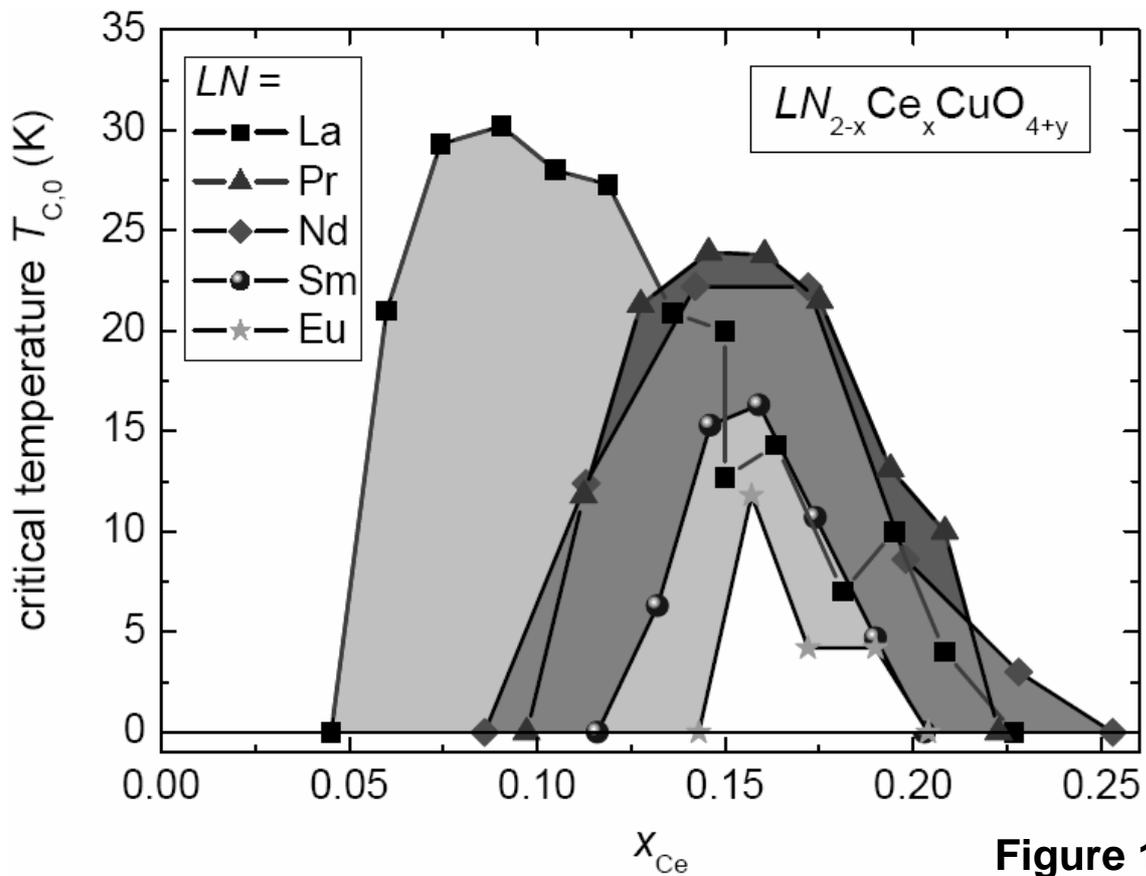

**Figure 1**

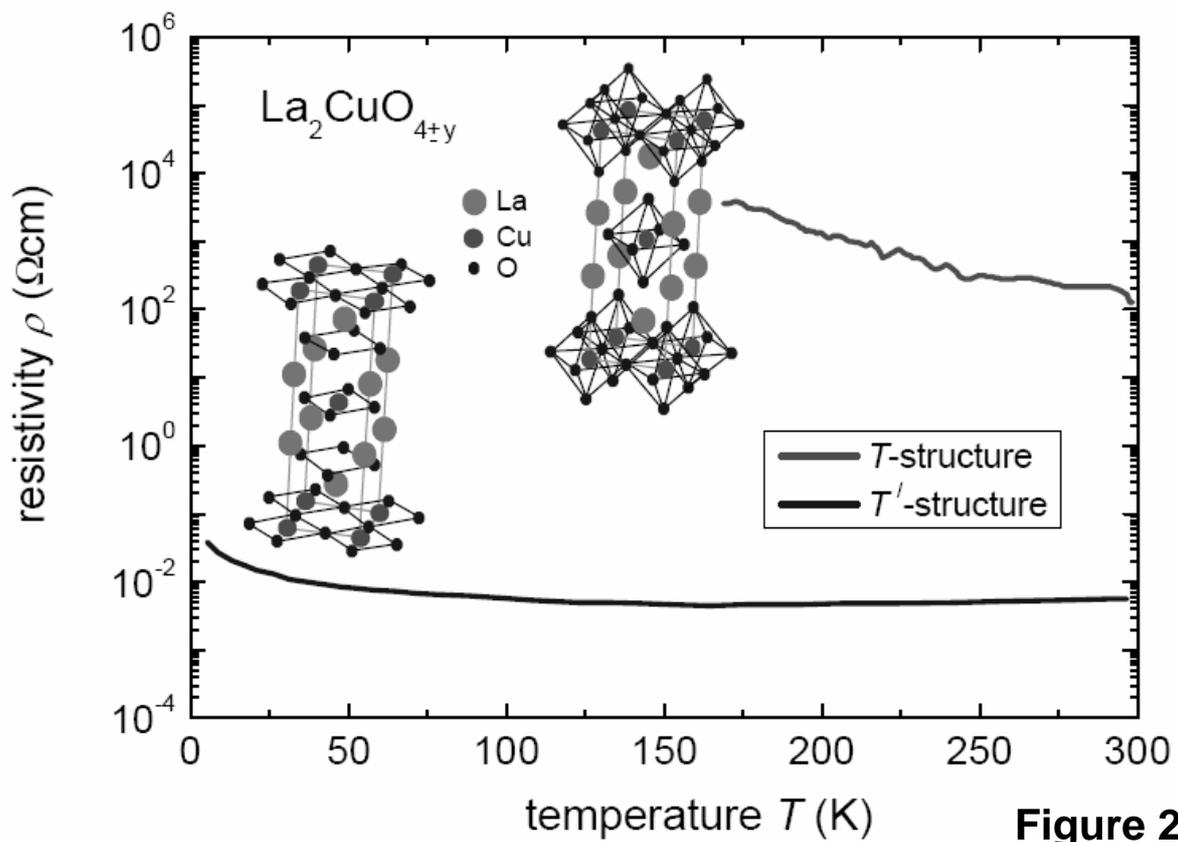

**Figure 2**

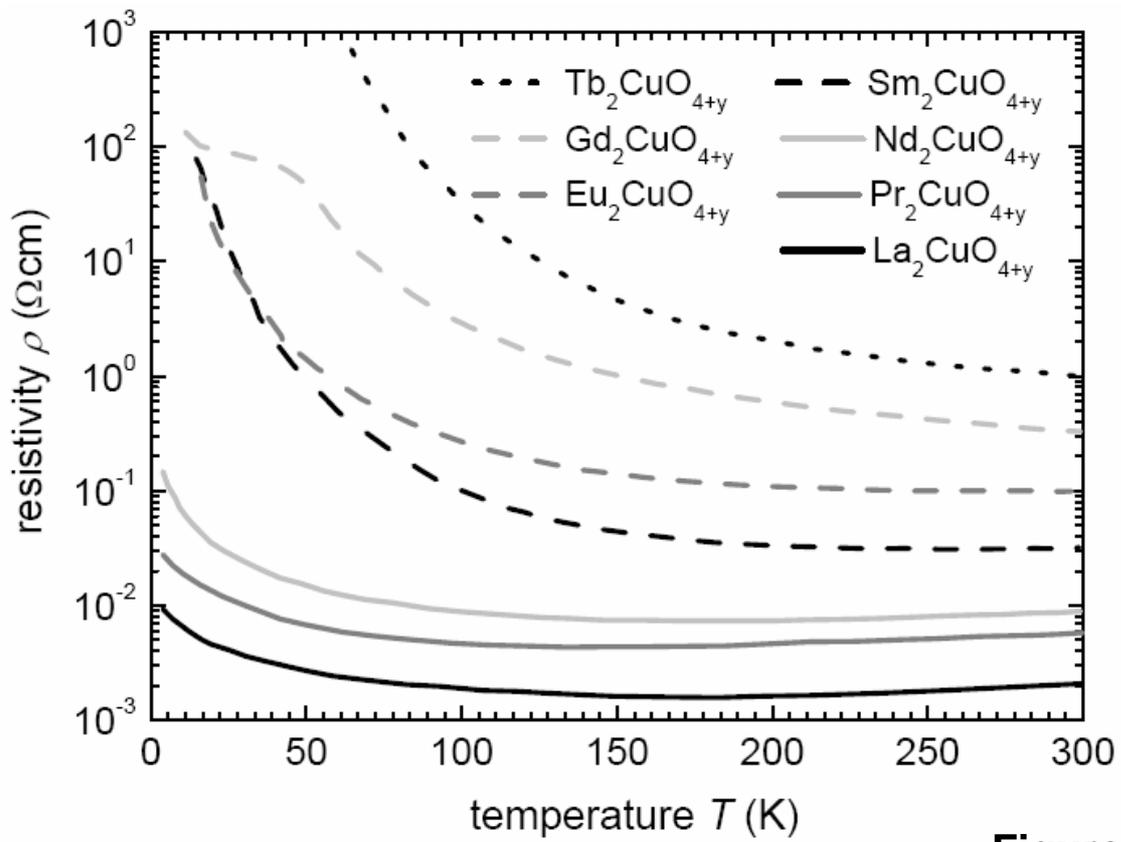

**Figure 3**

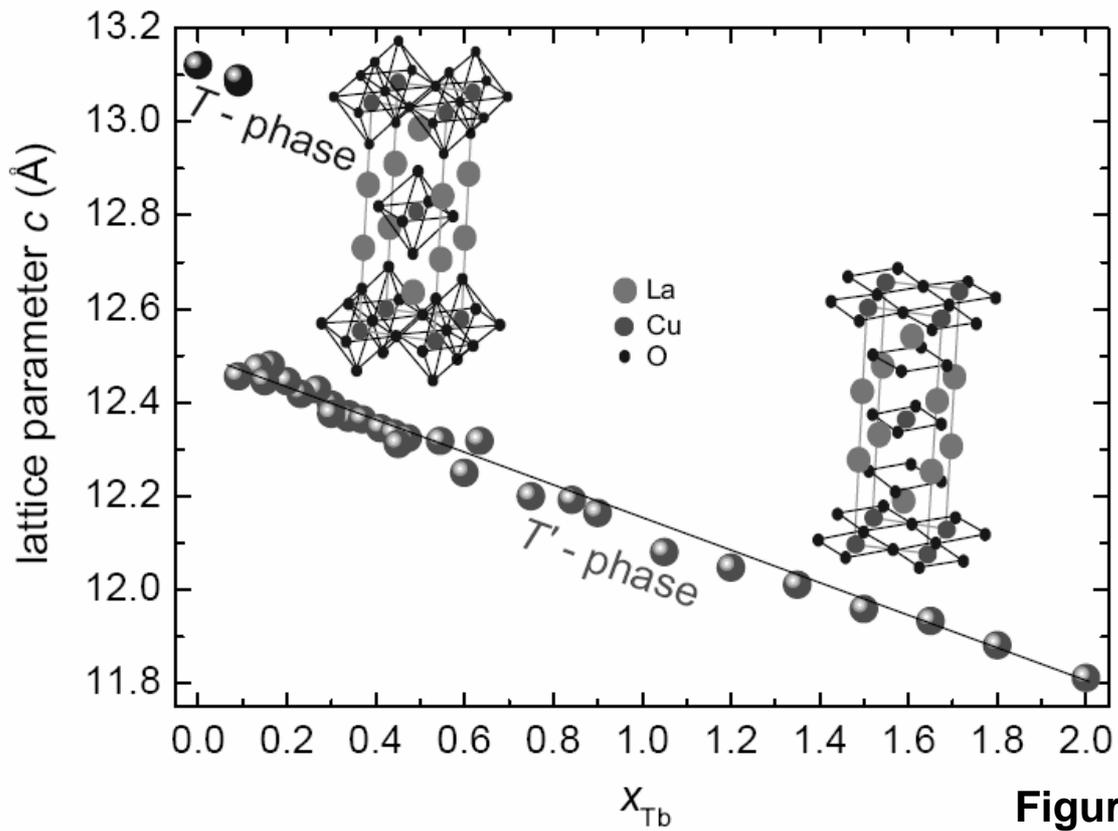

**Figure 4**

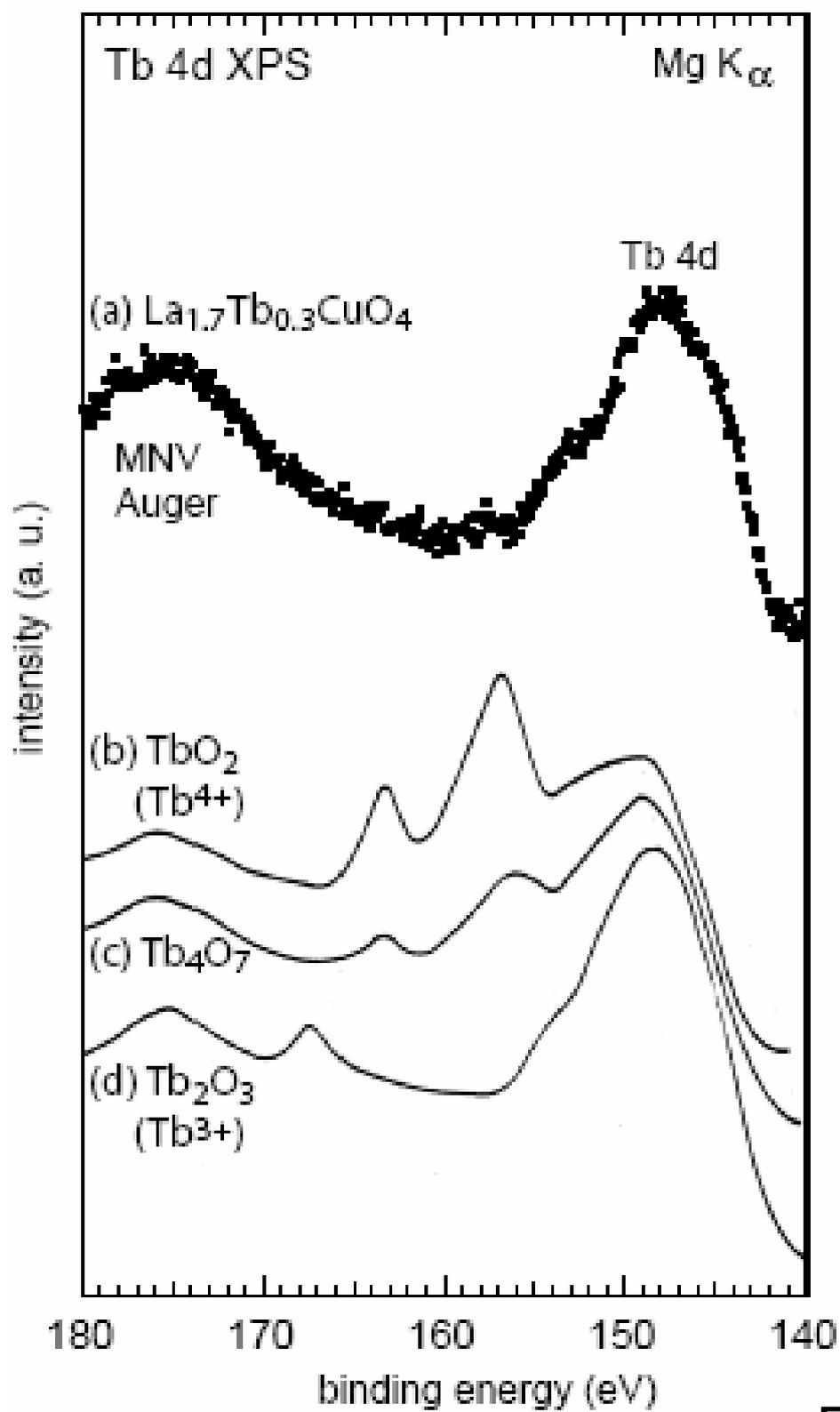

**Figure 5**

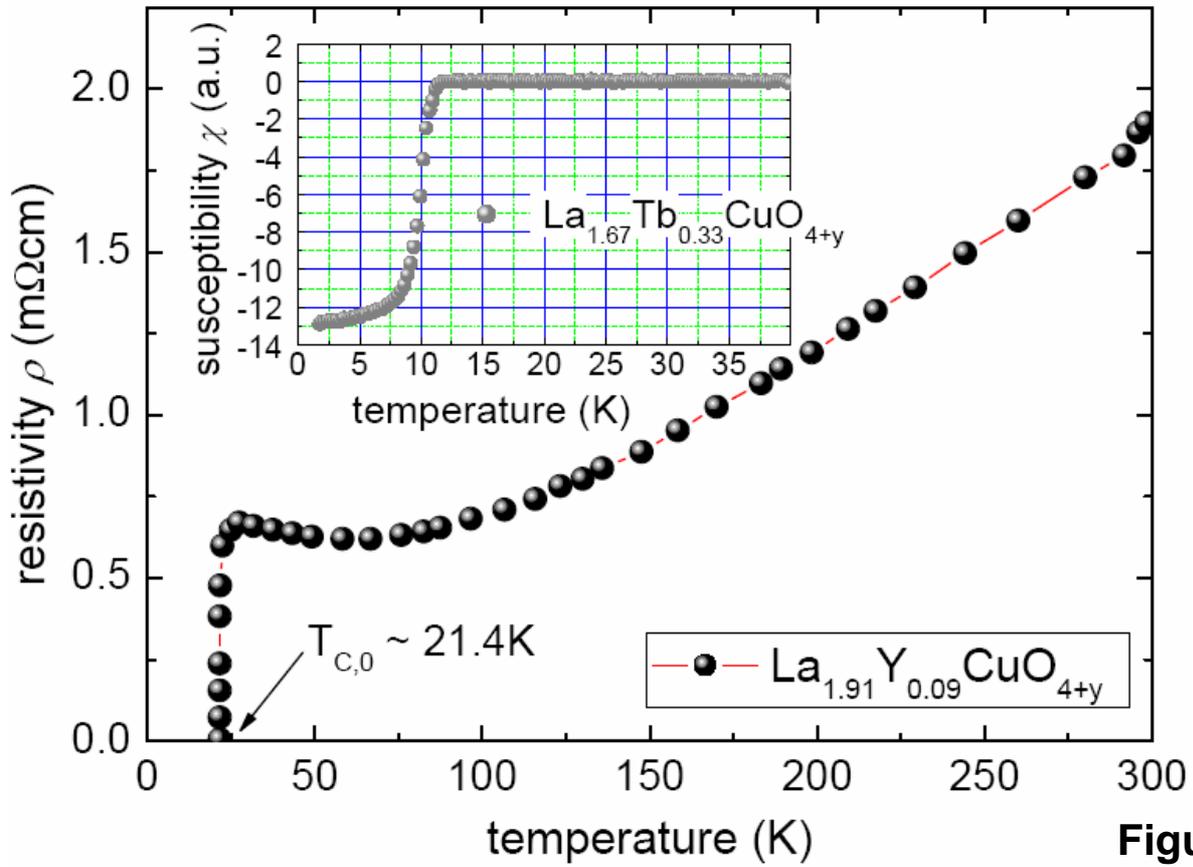

Figure 6

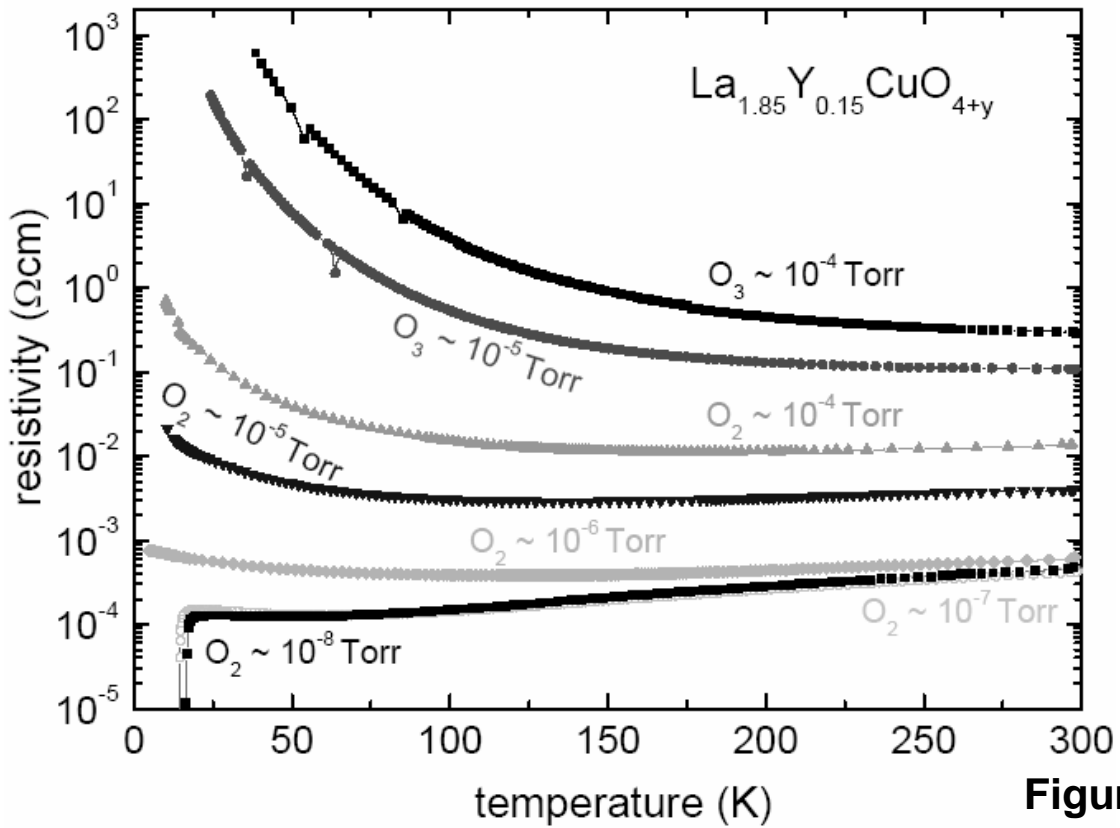

Figure 7